\documentstyle[aps,pre]{revtex}
\input{epsf}

\begin{document}
\title{Structure Factors and Their Distributions in Driven Two-Species Models}
\author{G. Korniss\footnote{Permanent address: Supercomputer
Computations Research Institute,\\ Florida State University, Tallahassee, 
Florida 32306-4052} and B. Schmittmann}
\address{Center for Stochastic Processes in Science and Engineering and
Department of Physics,\\ Virginia Polytechnic Institute and State University,
Blacksburg, Virginia 24061-0435}
\date{June 12, 1997}
\maketitle
\begin{abstract}
We study spatial correlations and structure factors in a three-state
stochastic lattice gas, consisting of holes and two oppositely ``charged''
species of particles, subject to an ``electric'' field at zero total charge.
The dynamics consists of two nearest-neighbor exchange processes, occuring
on different times scales, namely, particle-hole and particle-particle
exchanges. Using both, Langevin equations and Monte Carlo simulations, we
study the steady-state structure factors and correlation functions in the 
{\em disordered} phase, where density profiles are homogeneous. In contrast
to equilibrium systems, the {\em average} structure factors here show a
discontinuity singularity at the origin. The associated spatial correlation
functions exhibit intricate crossovers between exponential decays and power
laws of different kinds. The full probability distributions of the structure
factors are {\em universal} asymmetric exponential distributions.
\end{abstract}

\section{Introduction}

The study of correlations and structure factors provides a sensitive probe
into the characteristics of collective behavior in many-particle systems.
For example, in a system with short range microscopic interactions,
maintained in thermal equilibrium, spatial correlations will in general
decay exponentially. Long-range spatial correlations, characterized by power
laws, are observed only if the system is at a critical point. In contrast,
when such systems are driven into non-equilibrium steady states, long-range
correlations are often present in large regions of the phase diagram \cite
{liquids}. A model system in which such anomalous correlations are easily
observed and studied is the {\em driven} Ising lattice gas (the ``standard
model'') \cite{KLS,BS_RKPZ}. The drive (``external field'') biases particle
jumps along a specific lattice axis, thus forcing the system into a
non-equilibrium steady state. One of the most intriguing and unexpected
features of this system is the presence of long-range spatial correlations 
{\em at all temperatures }above criticality, due to the breakdown of the
traditional fluctuation-dissipation relations \cite{Kubo} in conjunction
with a conservation law for the particle density. In momentum space, the
structure factor develops a discontinuity singularity at the origin \cite
{ZHSL}.

In experiments, correlations are typically studied by photon, electron or
neutron beam scattering techniques. The scattering intensity is closely
related to the structure factor. Depending on the actual physical system,
this quantity is the Fourier transform of the ``density-density''
correlations, where, e.g., in a ferromagnetic system ``density'' stands for
the local magnetization. Even in the stationary case, i.e., when the
averages are not expected to be time-dependent, the densities themselves are
fluctuating quantities in both {\em space and time}. Thus, when data are
taken, it is crucial to compare the time scale of these fluctuations to the
duration of a typical ``snap-shot''. If the former is much shorter than the
sampling time interval, then even one measurement practically results in a
temporal average. In this case, the scattering intensity is a direct measure
of the {\em average} structure factor. In the opposite scenario each
individual snap shot appears as a random pattern of speckles. The collection
of many snap shots, however, represents the full distribution of the
fluctuating density-density products. This phenomenon has long been known in
laser scattering experiments and the statistical properties of the random
speckles are well established \cite{laser}. Using Monte Carlo simulations,
it is particularly easy to probe fluctuating quantities in terms of the
their distributions: each measurement corresponds to one configuration at a
certain instant of time so that there are no ``experimental'' difficulties
in achieving fine sampling. Since driven lattice gases exhibit generically
singular density-density correlations, a study of the associated
distributions is expected to be particularly interesting.

In this paper, we will focus on a generalization of the standard model,
similar in spirit to the one leading from the Ising model to spin-1 \cite
{BEG} or Potts \cite{Potts} models. Instead of just a single species of
particles, we now consider two (labelled as $+$'s and $-$'s) which are
driven in opposite directions, subject to periodic boundary conditions.
Empty lattice sites are referred to as holes. This generalization is
motivated by a variety of physical systems, ranging from fast ionic
conductors with several mobile ion species \cite{FIC} and water droplets in
microemulsions with distinct charges \cite{emuls} to gel electrophoresis 
\cite{electro} and traffic flow \cite{traffic}. For simplicity, we neglect
the usual Ising nearest neighbor interaction and retain ``only'' the
excluded volume constraint. The model thus corresponds to the
high-temperature, large-drive limit of a more complicated interacting system.

This multi-species model, in both one and two dimensions, has been studied
in detail \cite{traffic,SHZ,VZS,FG,KSZ,obc,Sandow}. In its simplest version,
particles are allowed to exchange with holes only. Monte Carlo simulations 
\cite{SHZ} in two dimensions and mean-field studies \cite{VZS,FG} show that
there is a transition, controlled by particle density and drive, from a
spatially homogeneous (disordered) phase to a charge segregated one, where
the excluded volume constraint leads to the mutual blocking of particles. In
this paper we extend our previous studies on structure factors\cite{KSZ1}
and their distributions \cite{KSZ2} to the more general case where the
particles are also allowed to exchange amongst themselves: we ``soften'' the
excluded volume constraint by allowing exchanges of nearest neighbor,
oppositely charged particles on a time scale $\gamma $ which is distinct
from that of the particle-hole exchange \cite{KSZ}. Here, the blocking
transition still occurs, as part of a more complex phase diagram \cite{KSZ}.
A particularly interesting case emerges when the two time scales are chosen
to be equal, i.e., $\gamma =1$: here, {\em equal} charges are completely
uncorrelated (up to trivial finite-size effects) while ``hidden'',
non-trivial correlations survive between {\em opposite} charges. We add that
this model in one dimension, at infinite drive and for arbitrary $\gamma $,
has been solved exactly by Godr\`{e}che and Sandow \cite{Sandow}.

We will focus on the disordered phase of the system, where we have a sound
analytic understanding of the dynamics in terms of Langevin equations. We
will study not only the averages but also the full distributions of the
steady-state structure factors, using Monte Carlo simulations and a
continuum field theory. Finding excellent agreement between our simulations
and analytic results, we can trace the characteristics of the distributions
back to the structure of the underlying Langevin equations. Given these
relations, measurements of structure factor distributions in real systems
should provide considerable information about the associated dynamics.

The paper is organized as follows. In the next section, we define the
microscopic model and give some details of the simulations. In Section III,
we introduce the underlying Langevin equations and calculate the average
structure factors, the corresponding spatial correlations, and the
theoretical distributions of the structure factors. In the final section, we
discuss our results and conclude with a brief summary.

\section{The Microscopic Model}

We consider a two dimensional fully periodic lattice with $L\times L$ sites,
each of which can be empty or occupied by a {\em single} particle. To
account for the presence of two species, we introduce two occupation numbers 
$n_{{\bf x}}^{+}$ and $n_{{\bf x}}^{-}$, with $n$ being 0 or 1, depending on
whether a positive or negative particle is present at site ${\bf x}$. The
excluded volume constraint implies \mbox{$\;n^{+}_{\bf x}n^{-}_{\bf x}=0\;$}%
, for any ${\bf x}$. To model the system at zero total charge, we choose 
\mbox{$\sum_{{\bf
x}}[n_{{\bf x}}^{+}-n_{{\bf x}}^{-}]=0$}, i.e., the average densities of
positive and negative particles are the same: 
\begin{equation}
\bar{\rho}=\frac{1}{L^{2}}\sum_{{\bf x}}n_{{\bf x}}^{+}=\frac{1}{L^{2}}\sum_{%
{\bf x}}n_{{\bf x}}^{-}\;,
\end{equation}
Since the dynamics conserves both densities separately, $\bar{\rho}$ is a
constant. In the absence of the drive, the two species of particles are
distinguished only by their label: both types hop randomly to
nearest-neighbor empty sites, with the same rate $\Gamma $. In addition,
nearest neighbor pairs of opposite charges exchange with a rate $\gamma
\Gamma $. The external drive is directed along a specific lattice axis,
labelled as the $+x_{\parallel }$-direction. Reminiscent of a uniform
``electric'' field $E$, it exponentially suppresses jumps {\em against} the
force. Specifically, during one Monte Carlo step $2L^{2}$ nearest-neighbor
bonds are selected at random. If a particle-hole pair is encountered, an
exchange takes place with probability 
\begin{equation}
W_{ph}=\Gamma \min \{1,\exp (qE\,\delta x_{\parallel })\}\;,  \label{phrate}
\end{equation}
where $q=\pm 1$ is the charge of the particle and $\delta x_{\parallel }=\pm
1,0$ is the change of the $x_{\parallel }$ coordinate of the particle due to
the jump. Similarly, if the neighboring sites are occupied by opposite
charges, a particle-particle exchange (or charge transfer) is attempted with
probability 
\begin{equation}
W_{pp}=\gamma \Gamma \min \{1,\exp (E\,\delta x_{\parallel })\}\;,
\label{pprate}
\end{equation}
where now $\delta x_{\parallel }$ is the change in the $x_{\parallel }$
coordinate of the {\em positive} particle due to the jump. Note that we do
not introduce a factor of $2$ in the exponential here, as one might have
expected for a real electric field. This choice leads to a simpler Langevin
equation without significantly affecting the phase diagram. Needless to say,
it is irrelevant whether exchange takes place or not, if both sites carry
identical content.

For our simulations, we set \mbox{$\Gamma=1$}, so the control parameters are 
$\bar{\rho}$, $E$ and $\gamma $. On lattices with $L$ ranging from $30$ to $%
100$, the system is initialized with random configurations of various
particle densities. Runs last from $2.5\times 10^{5}$ to $5\times 10^{5}$
MCS. The first $62500$ MCS are discarded to allow the system to settle into
steady state. Then, we measure the Fourier transforms of $n_{{\bf x}}^{\pm }$
every $125$ MCS, defining them in the usual way: 
\begin{equation}
n_{{\bf k}}^{\pm }=\sum_{{\bf x}}e^{-i\,{\bf kx}}\;n_{{\bf x}}^{\pm }\;\;.
\label{ftr_sim}
\end{equation}
In the following, we will investigate {\em equal-time }density-density
operators in momentum space, considering both their full distributions as
well as their (ensemble or time) averages. In the literature, the term
``structure factor'' typically refers to the averages, i.e., 
\begin{equation}
S^{\alpha \beta }({\bf k})\equiv \frac{1}{V}\langle n_{{\bf k}}^{\alpha }n_{-%
{\bf k}}^{\beta }\rangle \;,
\end{equation}
where $\alpha ,\beta =+,-$ ; ${\bf k}=\frac{2\pi }{L}(m_{\perp
},m_{\parallel })\neq {\bf 0}$ and $V=L^{2}$ is the volume. Occasionally,
especially when discussing the full distributions, we will use the word
``structure factor'' for the fluctuating two-point operator itself. In the
disordered phase, $S^{\alpha \beta }$ is the Fourier transform of the usual
equal-time correlation function 
\begin{equation}
G^{\alpha \beta }({\bf x})\equiv \langle n_{{\bf x}}^{\alpha }n_{_{{\bf 0}%
}}^{\beta }\rangle -\langle n_{{\bf x}}^{\alpha }\rangle \langle n_{_{{\bf 0}%
}}^{\beta }\rangle \;.  \label{def}
\end{equation}
Thus, if $G$ is even in ${\bf x}$, $S$ will be real, so that an imaginary
part of $S$ signals a part of $G$ which is odd in ${\bf x}$. By charge
symmetry, we expect $G^{++}=G^{--}$. Clearly, both must be even in ${\bf x}$%
, so that the associated $S$'s are real. On other hand, we have 
\begin{equation}
G^{+-}({\bf x},E)=G^{+-}(x_{\perp },-x_{\parallel },-E)  \label{g+-}
\end{equation}
in the presence of the drive so that $S^{+-}$ may have an imaginary part
(which must be odd in $E$). Finally, $G^{-+}({\bf x})=G^{+-}(-{\bf x})$
follows from (\ref{def}) by translation invariance. Turning to the full
distributions, these can be constructed from the time series of $\frac{n_{%
{\bf k}}^{+}n_{-{\bf k}}^{+}}{V}$, $\frac{\mbox{Re}[n_{{\bf k}}^{+}n_{-{\bf k%
}}^{-}]}{V}$ and $\frac{\mbox{Im}[n_{{\bf k}}^{+}n_{-{\bf k}}^{-}]}{V}$ in
the steady state. Exploiting symmetries again, we note that $\frac{n_{{\bf k}%
}^{+}n_{-{\bf k}}^{+}}{V}$ and $\frac{n_{{\bf k}}^{-}n_{-{\bf k}}^{-}}{V}$
are distributed identically, so that we need to consider only the former.
Further, only $\frac{n_{{\bf k}}^{+}n_{-{\bf k}}^{+}}{V}$ and $\frac{n_{{\bf %
k}}^{-}n_{-{\bf k}}^{-}}{V}$ are necessarily real, while $\frac{n_{{\bf k}%
}^{+}n_{-{\bf k}}^{-}}{V}$ will generically be complex.

We simulate systems with $\gamma $ ranging from $0$ to $1$. For small $%
\gamma $'s we choose $E$ and the density $\bar{\rho}$ in such a way that the
system is in the homogeneous phase. For larger $\gamma $'s ($\gamma >\gamma
_{c}\simeq 0.62$) the charge exchange mechanism suppresses the ordered phase
entirely \cite{KSZ} so we can pick arbitrarily large fields at any density.
A particularly interesting case occurs for $\gamma =1$. Here, the rates for
particle-hole and particle-particle exchanges become equal, i.e., $%
W_{pp}=W_{ph}$, so that a positive (negative) particle can no longer
distinguish a negative (positive) one from a hole. Thus, a positive
(negative) particle experiences biased diffusion, slowed only by encounters
with other positive (negative) particles, just as in the case of a single,
non-interacting species, whose steady state probability distribution of
configurations (i.e., the steady state solution of the associated master
equation) is exactly known to be uniform \cite{Spitz}. For our case, this
implies that the {\em marginal} distribution of the occupation numbers of
one species is uniform, i.e., $P\left[ \{n_{{\bf x}}^{\pm }\}\right]
=\sum_{\{n_{{\bf x}}^{\mp }\}}P\left[ \{n_{{\bf x}}^{+},n_{{\bf x}%
}^{-}\}\right] \propto 1.$ Thus, we expect questions regarding only one
species of particles to have trivial answers, e.g. $G^{++}({\bf x})$ must
vanish for ${\bf x}\neq {\bf 0}$ in an infinite system or yield the
finite-size fluctuations in a finite one. On the other hand, the two-point
function between opposite charges can display interesting structures, e.g.
long-range correlations, as a result of the full distribution $P\left[ \{n_{%
{\bf x}}^{+},n_{{\bf x}}^{-}\}\right] $ not being uniform. We note briefly
that a completely ``flat'' steady state, $P\left[ \{n_{{\bf x}}^{+},n_{{\bf x%
}}^{-}\}\right] \propto 1$, is obtained for $\gamma =2$, as in the
one-dimensional version of our model \cite{Sandow}.

In Fig. \ref{fig1}, we present the results for the three independent $S$'s
found in the $100\times 100$ system at a small value of $\gamma $ and note
the discontinuity singularity of these objects at the origin. In Fig. \ref
{fig2}, we show the same quantities for $\gamma =1$ and draw special
attention to the fact that, while $S^{++}$ does not depend on ${\bf k}$ at
all, $S^{+-}$ exhibits a highly nontrivial ${\bf k}$-dependence. Fig. \ref
{fig3} and \ref{fig4} present the structure factor distributions for the
smallest longitudinal and transverse wave vectors, respectively. Before
discussing the data in detail, we will first present the theoretical
framework within which they can be understood. In particular, we will focus
on two points, namely first, the emergence of discontinuity singularities in
the structure factors at ${\bf k=0}$, and their consequences for long-range
correlations in real space, and second, the origin of the asymmetric
exponential form of the distributions. This will then be followed by a
comparison between our theoretical predictions and the simulations.

\section{Coarse-grained Description}

To extract the behavior at large distances (or small $k$ in momentum space),
a continuum field theory for the slow variables of the model is most
appropriate. To find such a description, we must (i) identify the slow
variables of the theory, and (ii) obtain a set of equations of motion for
these quantities, corresponding to a coarse-grained version of the
microscopic dynamics. For generality, we consider the $d$-dimensional case
when ${\bf x}_{\parallel }$ is directed along the electric field and ${\bf x}%
_{\perp }$ is in the ($d-1$)-dimensional subspace, perpendicular to the
field. Time is denoted by $t$. Then, the slow variables are easily
identified as the conserved densities, $\rho ^{\pm }({\bf x},t)$, of the two
species. The most systematic way to arrive at their equations of motion is
to perform an $\Omega $-expansion \cite{VK,PhD}: After partitioning the
whole system into sufficiently large blocks of size $\Omega $, one splits
the particle densities associated with the block centered at ${\bf {x}}$
into a macroscopic part ($\rho ^{\pm }$) and a fluctuating one ($\chi ^{\pm
} $) : 
\begin{equation}
\frac 1\Omega \sum_{{\bf x}^{^{\prime }}\,\epsilon \,\mbox{b}({\bf x})}n_{%
{\bf x}^{^{\prime }}}^{\pm }=\rho ^{\pm }({\bf x},t)+\Omega ^{-1/2}\chi
^{\pm }({\bf x},t)\;.  \label{omega}
\end{equation}
This decomposition is inserted into the microscopic master equation,
followed by a systematic expansion in $\Omega $. At leading order, we find a
set of mean-field equations of motion for the local densities which reads,
after taking a naive continuum limit: 
\begin{equation}
\partial _t\rho ^{\pm }=-\mbox{\boldmath $\nabla$}{\bf \Gamma }\left\{ [\rho
^{\pm }\stackrel{\leftrightarrow }{\mbox{\boldmath$\nabla$}}(1-\rho
^{+}-\rho ^{-})\pm \varepsilon \hat{{\bf x}}_{\parallel }\rho ^{\pm }(1-\rho
^{+}-\rho ^{-})]+\gamma [\rho ^{\pm }\stackrel{\leftrightarrow }{%
\mbox{\boldmath$\nabla$}}\rho ^{\mp }\pm \varepsilon \hat{{\bf x}}%
_{\parallel }\rho ^{\pm }\rho ^{\mp }]\right\} ,  \label{meaf}
\end{equation}
where 
\begin{equation}
{\bf \Gamma }=\left( \matrix{ {\bf \Gamma}_{\perp} & {\bf 0} \cr {\bf 0} &
\Gamma_{\parallel} }\right)  \label{diffmatr}
\end{equation}
is the diffusion-matrix. ${\bf \Gamma }_{\perp }$ is diagonal and isotropic
in the ($d-1$)-dimensional subspace, thus characterized by a number $\Gamma
_{\perp }$. $\stackrel{\leftrightarrow }{\mbox{\boldmath $\nabla$}}$ is the
asymmetric gradient operator, acting on any two functions $f$ and $g$
according to 
\mbox{$f\stackrel{\leftrightarrow }{\mbox{\boldmath $\nabla$}}g=
f\mbox{\boldmath $\nabla$}g-g\mbox{\boldmath $\nabla$}f$}. $\varepsilon $ is
the coarse-grained bias and $\hat{{\bf x}}_{\parallel }$ is the unit vector
along the $x_{\parallel }$ direction. Note that at the mean-field level we
also obtain explicit expressions for the diffusion matrix and the bias \cite
{PhD}: \mbox{$\Gamma_{\perp}=1$}, \mbox{$\Gamma_{\parallel}=(1+e^{-|E|})/2$}
and \mbox{$\varepsilon=2\tanh(E/2)$}. Of course, these may be modified by
renormalization.

The continuity equation (\ref{meaf}) admits both homogeneous and
inhomogeneous $t$-independent solutions, associated with the disordered and
the blocked phases. The former is our focus here. To ease comparison with
simulation data, we choose equal densities for both species: $\rho ^{\pm }(%
{\bf x},t)=\bar{\rho}$ . This solution describes the steady state at the
mean-field level.

At the next order in the $\Omega $-expansion, we find a Fokker-Planck
equation for the fluctuating part, $\chi ^{\pm }$. For our purposes, the
equivalent Langevin equation is more transparent. At this order, its
deterministic part is linear and the (conserved) noise is Gaussian. After
defining the ``reduced'' average density $\tilde{\rho}\equiv (1-\gamma )\bar{%
\rho}$ and $\delta \equiv (3-\gamma )/(1-\gamma )$, and focusing on the
fluctuations about the homogeneous phase, the result is: 
\begin{equation}
\partial _{t}\chi ^{\alpha }({\bf x},t)={\cal L}^{\alpha \beta }(%
\mbox{\boldmath 
$\nabla$})\chi ^{\beta }({\bf x},t)-\mbox{\boldmath $\nabla \eta $}^{\alpha
}({\bf x},t)\;,  \label{rsl}
\end{equation}
where the drift matrix is given by 
\begin{equation}
({\cal L}^{\alpha \beta }(\mbox{\boldmath $\nabla$}))=\left( \matrix{
(1-\tilde{\rho})\mbox{\boldmath $\nabla\Gamma\nabla$} -
(1-\delta\tilde{\rho}) \varepsilon\Gamma_{\parallel}\partial_{\parallel} &
\tilde{\rho}\mbox{\boldmath $\nabla\Gamma\nabla$} + \tilde{\rho}
\varepsilon\Gamma_{\parallel}\partial_{\parallel} \cr
\tilde{\rho}\mbox{\boldmath $\nabla\Gamma\nabla$} - \tilde{\rho}
\varepsilon\Gamma_{\parallel}\partial_{\parallel} &
(1-\tilde{\rho})\mbox{\boldmath $\nabla\Gamma\nabla$} +
(1-\delta\tilde{\rho}) \varepsilon\Gamma_{\parallel}\partial_{\parallel} }%
\right)  \label{det_matr}
\end{equation}
and summation over repeated indices is implied in (\ref{rsl}) and in the
following. The $\mbox{\boldmath $\eta $}^{\pm }({\bf x},t)$ are Gaussian
white noise terms, with average and second moment: 
\begin{eqnarray}
\langle \eta _{i}^{\alpha }({\bf x},t)\rangle & = & 0, \nonumber \\ 
\;\;\langle \eta _{i}^{\alpha }({\bf x},t)\eta _{j}^{\beta }({\bf x}^{\prime
},t^{\prime })\rangle & = & 2\sigma _{ij}^{\alpha \beta }\delta ({\bf x}-%
{\bf x}^{\prime })\delta (t-t^{\prime })\;, \label{noise}
\end{eqnarray}
where $\alpha ,\beta =+,-$ ; $i,j=1,2,\ldots d$. Due to the bias, the noise
matrices $(\sigma _{ij}^{\alpha \beta })=\mbox{\boldmath $\sigma$}^{\alpha
\beta }$ are diagonal but not proportional to the unit matrix: 
\begin{equation}
\mbox{\boldmath $\sigma$}^{\alpha \beta }=\left( \matrix{
\mbox{\boldmath$\sigma$}^{\alpha\beta}_{\perp} & {\bf 0} \cr {\bf 0} &
\sigma^{\alpha\beta}_{\parallel} }\right) \;.  \label{noisematr}
\end{equation}
Note that $\mbox{\boldmath $\sigma$}^{\alpha \beta }$ is symmetric and due
to charge symmetry, we also have $\mbox{\boldmath $\sigma$}^{++}=%
\mbox{\boldmath $\sigma$}^{--}$. Similar to ${\bf \Gamma }_{\perp }$, $%
\mbox{\boldmath $\sigma$}_{\perp }^{\alpha \beta }$ is diagonal and
isotropic in the ($d-1$)-dimensional subspace, characterized by a number $%
\sigma _{\perp }^{\alpha \beta }$. In the absence of the drive, our model
reduces to an equilibrium system, so that the fluctuation dissipation
theorem (FDT) holds. In our case, this guarantees $\mbox{\boldmath $\sigma$}%
^{\alpha \beta }\propto {\bf \Gamma }$, or, more specifically, 
\begin{eqnarray}
\mbox{\boldmath $\sigma$}^{++} & = & \left[ \bar{\rho}(1-2\bar{\rho})+\gamma 
\bar{\rho}^{2}\right] {\bf \Gamma } \nonumber \\
\mbox{\boldmath $\sigma$}^{+-} & = & \left[ -\gamma \bar{\rho}^{2}\right] 
{\bf \Gamma } \label{FDT}
\end{eqnarray}
\cite{PhD}. However, when driven, this proportionality does not hold in
generic ranges of $\gamma $ and $\bar{\rho}$, in that the diffusion and
noise matrices are renormalized differently by the drive $\varepsilon $,
similar to the situation in the driven single species case \cite{JS_LC}.
Finally, we point out that there is a correlation between $%
\mbox{\boldmath
$\eta $}^{+}$ and $\mbox{\boldmath $\eta $}^{-}$ due to the fact that charge
exchange is allowed. This effect is captured by the matrix $%
\mbox{\boldmath
$\sigma$}^{+-}$ which is expected to be proportional to $\gamma $ and
negative definite for non-zero drive as well.

\subsection{Steady-state structure factors}

Eqns. (\ref{rsl}-\ref{noisematr}) are linear equations which are easily
solved in Fourier space. Introducing the Fourier components for the
fluctuations 
\begin{equation}
\chi ^{\pm }({\bf k},\omega )=\int dtd^{d}x\,\chi ^{\pm }({\bf x}%
,t)\,e^{-i(\omega t+{\bf kx})}\;,  \label{ftrdef}
\end{equation}
and similar ones for the noise, so that 
\begin{eqnarray}
\langle \eta _{i}^{\alpha }({\bf k},\omega )\rangle & = & 0 \nonumber \\ 
\langle \eta _{i}^{\alpha }({\bf k},\omega )\eta _{j}^{\beta }({\bf k}%
^{\prime },\omega ^{\prime })\rangle & = & 2\sigma _{ij}^{\alpha \beta
}\left[ (2\pi )^{d+1}\delta ({\bf k}+{\bf k}^{\prime })\delta (\omega
+\omega ^{\prime })\right] \label{ksnoise}
\end{eqnarray}
the solution to (\ref{rsl}) is simply: 
\begin{equation}
\chi ^{\alpha }({\bf k},\omega )=(L^{-1})^{\alpha \beta }\;i{\bf k}%
\mbox{\boldmath $\eta $}^{\beta }({\bf k},\omega )\;,  \label{inv_kol}
\end{equation}
where 
\begin{equation}
L^{\alpha \beta }({\bf k},\omega )\equiv {\cal L}^{\alpha \beta }(i{\bf k}%
)-i\omega \delta ^{\alpha \beta }.  \label{Ldef}
\end{equation}
Note that, in ${\bf k}$ space, $\left( {\cal L}^{++},{\cal L}^{--}\right) $
and $\left( {\cal L}^{+-},{\cal L}^{-+}\right) $ are complex conjugate pairs.

Not surprisingly, $\langle \chi ^{\pm }({\bf k},\omega )\rangle =0$,
consistent with the decomposition (\ref{omega}). The two-point correlations
of $\chi ^{\pm }({\bf k},\omega )$ are just the {\em dynamic} structure
factors, defined as 
\begin{equation}
S^{\alpha \beta }({\bf k},\omega )\left[ (2\pi )^{d+1}\delta ({\bf k}+{\bf k}%
^{\prime })\delta (\omega +\omega ^{\prime })\right] \equiv \langle \chi
^{\alpha }({\bf k},\omega )\chi ^{\beta }({\bf k}^{\prime },\omega ^{\prime
})\rangle \;\text{.}  \label{dyn_str_def}
\end{equation}
Using (\ref{inv_kol}) and (\ref{ksnoise}), the two independent $S$'s follow: 
\begin{eqnarray}
S^{++}({\bf k},\omega ) &=&\frac{2\,{\bf k}\mbox{\boldmath $\sigma$}^{++}%
{\bf k}}{\mid \det (L)\mid ^{2}}\,\left( \mid L^{--}\mid ^{2}+\mid
L^{+-}\mid ^{2}\right) -\frac{2\,{\bf k}\mbox{\boldmath $\sigma$}^{+-}{\bf k}%
}{\mid \det (L)\mid ^{2}}\,2\mbox{Re}\{L^{--}L^{-+}\}  \nonumber \\
S^{+-}({\bf k},\omega ) &=&-\frac{2\,{\bf k}\mbox{\boldmath $\sigma$}^{++}%
{\bf k}}{\mid \det (L)\mid ^{2}}\,L^{+-}\left( {(L^{++})}^{*}+L^{--}\right) +%
\frac{2\,{\bf k}\mbox{\boldmath $\sigma$}^{+-}{\bf k}}{\mid \det (L)\mid ^{2}%
}\,\left( {(L^{++})}^{*}L^{--}+{(L^{+-})}^{2}\right) .  \label{dyn_str}
\end{eqnarray}
To compare directly with simulations, we need the steady-state structure
factors 
\begin{equation}
S^{\alpha \beta }({\bf k})\left[ (2\pi )^{d}\delta ({\bf k}+{\bf k}^{\prime
})\right] \equiv \langle \chi ^{\alpha }({\bf k},t)\chi ^{\beta }({\bf k}%
^{\prime },t)\rangle \;,  \label{str_def}
\end{equation}
which are easily obtained from (\ref{dyn_str}) by an integration over $%
\omega $, using the residue theorem and noting that the two zeros of %
\mbox{$\det(L)$} simply correspond to the two stable eigenvalues of ${\cal L}
$: 
\begin{equation}
\omega _{1,2}=-i\frac{\mbox{Tr}({\cal L})}{2}\pm \sqrt{\det ({\cal L}%
)-\left( \frac{\mbox{Tr}({\cal L})}{2}\right) ^{2}}\;.  \label{residues}
\end{equation}
To ensure that the system is within the linear stability region of the
disordered phase, we must have $\mbox{Im}\,\omega_{1,2}>0$. 
Since $-\frac{1}{2}\mbox{Tr}({\cal L})=(1-\tilde{\rho})%
{\bf k\Gamma k}$ is automatically positive definite, we only require $\det (%
{\cal L})>0$ for all ${\bf k}\neq 0$. Collecting, we find: 
\begin{eqnarray}
S^{++}({\bf k}) &=&\frac{{\bf k}\mbox{\boldmath $\sigma$}^{++}{\bf k}}{-%
\frac{1}{2}\mbox{Tr}({\cal L})}\,\frac{\mid {\cal L}^{--}\mid ^{2}}{\det (%
{\cal L})}-\frac{{\bf k}\mbox{\boldmath $\sigma$}^{+-}{\bf k}}{-\frac{1}{2}%
\mbox{Tr}({\cal L})}\,\frac{\mbox{Re}\{{\cal L}^{--}{\cal L}^{-+}\}}{\det (%
{\cal L})}  \nonumber \\
S^{+-}({\bf k}) &=&-\frac{{\bf k}\mbox{\boldmath $\sigma$}^{++}{\bf k}}{-%
\frac{1}{2}\mbox{Tr}({\cal L})}\,\frac{{\cal L}^{--}{\cal L}^{+-}}{\det (%
{\cal L})}+\frac{{\bf k}\mbox{\boldmath $\sigma$}^{+-}{\bf k}}{-\frac{1}{2}%
\mbox{Tr}({\cal L})}\,\frac{{\cal L}^{--}\mbox{Re}\{{\cal L}^{--}\}+i{\cal L}%
^{+-}\mbox{Im}\{{\cal L}^{+-}\}}{\det ({\cal L})}\;  \label{steady_str}
\end{eqnarray}
so that, with the help of (\ref{det_matr}), we finally obtain: 
\begin{eqnarray}
S^{++}({\bf k}) &=&\frac{(1-\tilde{\rho})}{(1-2\tilde{\rho})}\,\frac{{\bf k}%
\mbox{\boldmath $\sigma$}^{++}{\bf k}}{{\bf k\Gamma k}}\;\frac{({\bf k\Gamma
k})^{2}+\frac{(1-\delta \tilde{\rho})^{2}}{(1-\tilde{\rho})^{2}}\varepsilon
^{2}\Gamma _{\parallel }^{2}k_{\parallel }^{2}}{({\bf k\Gamma k}%
)^{2}+4m^{2}\Gamma _{\parallel }k_{\parallel }^{2}}  \nonumber \\
&-&\frac{\tilde{\rho}}{(1-2\tilde{\rho})}\,\frac{{\bf k}%
\mbox{\boldmath
$\sigma$}^{+-}{\bf k}}{{\bf k\Gamma k}}\;\frac{({\bf k\Gamma k})^{2}+\frac{%
(1-\delta \tilde{\rho})}{(1-\tilde{\rho})}\varepsilon ^{2}\Gamma _{\parallel
}^{2}k_{\parallel }^{2}}{({\bf k\Gamma k})^{2}+4m^{2}\Gamma _{\parallel
}k_{\parallel }^{2}}  \nonumber \\
\mbox{Re}\{S^{+-}({\bf k})\} &=&-\frac{\tilde{\rho}}{(1-2\tilde{\rho})}\,%
\frac{{\bf k}\mbox{\boldmath $\sigma$}^{++}{\bf k}}{{\bf k\Gamma k}}\;\frac{(%
{\bf k\Gamma k})^{2}-\frac{(1-\delta \tilde{\rho})}{(1-\tilde{\rho})}%
\varepsilon ^{2}\Gamma _{\parallel }^{2}k_{\parallel }^{2}}{({\bf k\Gamma k}%
)^{2}+4m^{2}\Gamma _{\parallel }k_{\parallel }^{2}}  \nonumber \\
&+&\frac{(1-\tilde{\rho})}{(1-2\tilde{\rho})}\,\frac{{\bf k}%
\mbox{\boldmath
$\sigma$}^{+-}{\bf k}}{{\bf k\Gamma k}}\;\frac{({\bf k\Gamma k})^{2}-\frac{{%
\tilde{\rho}}^{2}}{(1-\tilde{\rho})^{2}}\varepsilon ^{2}\Gamma _{\parallel
}^{2}k_{\parallel }^{2}}{({\bf k\Gamma k})^{2}+4m^{2}\Gamma _{\parallel
}k_{\parallel }^{2}}  \label{exp_str} \\
\mbox{Im}\{S^{+-}({\bf k})\} &=&\frac{\tilde{\rho}(2-(1+\delta )\tilde{\rho})%
}{(1-\tilde{\rho})(1-2\tilde{\rho})}\;\frac{({\bf k}%
\mbox{\boldmath
$\sigma$}^{++}{\bf k})\,\varepsilon \Gamma _{\parallel }k_{\parallel }}{(%
{\bf k\Gamma k})^{2}+4m^{2}\Gamma _{\parallel }k_{\parallel }^{2}}  \nonumber
\\
&-&\frac{(1-\tilde{\rho})(1-\delta \tilde{\rho})+{\tilde{\rho}}^{2}}{(1-%
\tilde{\rho})(1-2\tilde{\rho})}\;\frac{({\bf k}\mbox{\boldmath $\sigma$}^{+-}%
{\bf k})\,\varepsilon \Gamma _{\parallel }k_{\parallel }}{({\bf k\Gamma k}%
)^{2}+4m^{2}\Gamma _{\parallel }k_{\parallel }^{2}}\;,  \nonumber
\end{eqnarray}
To simplify the notation, we have defined a ``mass'' (in the field theory
sense), $m$ , via 
\begin{equation}
4m^{2}\equiv \frac{(1-\delta \tilde{\rho})^{2}-\tilde{\rho}^{2}}{1-2\tilde{%
\rho}}\;\varepsilon ^{2}\Gamma _{\parallel }=\frac{(1-2\bar{\rho}%
)(1-(2-\gamma )2\bar{\rho})}{(1-(1-\gamma )2\bar{\rho})}\;\varepsilon
^{2}\Gamma _{\parallel }  \label{mdef}
\end{equation}
Its role is to mark the linear stability boundary, which, in the limit $%
\varepsilon L\rightarrow \infty $, is given precisely by $m^{2}=0$ .
Otherwise, for {\em finite }$\varepsilon L$, the system does not reach the
stability limit as long as $(\varepsilon L/2\pi )^{2}<(1-(1-\gamma )2\bar{%
\rho})/(1-2\bar{\rho})((2-\gamma )2\bar{\rho}-1)$ is satisfied \cite{KSZ}.
Thus, it is sufficient to impose $m^{2}>0$, i.e., $\bar{\rho}<\frac{1}{%
2(2-\gamma )}$, to keep the system in the homogeneous phase.

Similar to the driven lattice gas \cite{KLS,BS_RKPZ} and the two-species
model studied earlier \cite{KSZ2}, these structure factors are all singular
at the origin. The singularity takes the form of a discontinuity, either in
the function itself or one of its derivatives. In particular, both $S^{++}$
and $\mbox{Re}\{S^{+-}\}$ are discontinuous, so that the ratios 
\begin{equation}
\frac{\lim_{k_{\parallel }\rightarrow 0}S^{++}({\bf 0},k_{\parallel })}{%
\lim_{{\bf k}_{\perp }\rightarrow {\bf 0}}S^{++}({\bf k}_{\perp },0)}=\frac{%
1-2\tilde{\rho}}{(1-\delta \tilde{\rho})^{2}-\tilde{\rho}^{2}}\;\frac{%
(1-\delta \tilde{\rho})^{2}}{(1-\tilde{\rho})^{2}}\;\frac{\frac{\sigma
_{\parallel }^{++}}{\Gamma _{\parallel }}-\frac{\tilde{\rho}}{1-\delta 
\tilde{\rho}}\;\frac{\sigma _{\parallel }^{+-}}{\Gamma _{\parallel }}}{\frac{%
\sigma _{\perp }^{++}}{\Gamma _{\perp }}-\frac{\tilde{\rho}}{1-\tilde{\rho}}%
\;\frac{\sigma _{\perp }^{+-}}{\Gamma _{\perp }}}  \label{S++_disc}
\end{equation}
and 
\begin{equation}
\frac{\lim_{k_{\parallel }\rightarrow 0}\mbox{Re}\{S^{+-}({\bf 0}%
,k_{\parallel })\}}{\lim_{{\bf k}_{\perp }\rightarrow {\bf 0}}\mbox{Re}%
\{S^{+-}({\bf k}_{\perp },0)\}}=-\frac{1-2\tilde{\rho}}{(1-\delta \tilde{\rho%
})^{2}-\tilde{\rho}^{2}}\;\frac{1-\delta \tilde{\rho}}{1-\tilde{\rho}}\;%
\frac{\frac{\sigma _{\parallel }^{++}}{\Gamma _{\parallel }}-\frac{\tilde{%
\rho}}{1-\delta \tilde{\rho}}\;\frac{\sigma _{\parallel }^{+-}}{\Gamma
_{\parallel }}}{\frac{\sigma _{\perp }^{++}}{\Gamma _{\perp }}-\frac{1-%
\tilde{\rho}}{\tilde{\rho}}\;\frac{\sigma _{\perp }^{+-}}{\Gamma _{\perp }}}%
\;.  \label{S+-_disc}
\end{equation}
are in general different from unity. In contrast, $\mbox{Im}\{S^{+-}({\bf k}%
)\}$ vanishes for ${\bf k}\rightarrow {\bf 0}$ in any direction. Here,
discontinuities occur in higher derivatives. Unlike in the driven Ising
lattice gas, these singularities do not simply originate in the generic
FDT-breaking relation 
\mbox{$\frac{\sigma _{\parallel }^{\alpha \beta }}{\Gamma _{\parallel }}\neq
\frac{\sigma
_{\perp }^{\alpha \beta }}{\Gamma _{\perp }}$}, but also in the specifics of
this particular driven system, reflected in the first factor on the right
hand side of (\ref{S++_disc},\ref{S+-_disc}). It is a monotonically
increasing function of $\bar{\rho}$, reaching $\infty $ at $\bar{\rho}=\frac{%
1}{2(2-\gamma )}$. As a result, the amplitudes of the discontinuities
diverge as the system approaches the stability limit of the homogeneous
phase.

\subsection{Equal-time spatial correlations}

The equal-time correlation functions $G^{\alpha \beta }({\bf x})\equiv
\langle \chi ^{\alpha }({\bf x}^{\prime }+{\bf x},t)\chi ^{\beta }({\bf x}%
^{\prime },t)\rangle $ are just the Fourier transforms of the structure
factors, 
\begin{equation}
G^{\alpha \beta }({\bf x})=\int \frac{d^{d}k}{(2\pi )^{d}}\,S^{\alpha \beta
}({\bf k})e^{i{\bf kx}}\;.  \label{spcorr_def}
\end{equation}
independent of ${\bf x}^{\prime }$ by virtue of translational invariance. To
simplify the transforms, we introduce some changes in notation. First, we
rescale the lengths and momenta ${\bf x}_{\perp }\rightarrow {\bf x}_{\perp
}/\Gamma _{\perp }^{\frac{1}{2}}$, $x_{\parallel }\rightarrow x_{\parallel
}/\Gamma _{\parallel }^{\frac{1}{2}}$; ${\bf k}_{\perp }\rightarrow \Gamma
_{\perp }^{\frac{1}{2}}{\bf k}_{\perp }$ ,$\;k_{\parallel }\rightarrow
\Gamma _{\parallel }^{\frac{1}{2}}k_{\parallel }$ so that ${\bf \Gamma }$
becomes the unit matrix. Further, we let $\sigma _{\perp }^{\alpha \beta
}\rightarrow \sigma _{\perp }^{\alpha \beta }/\Gamma _{\perp }$ ,$\;\sigma
_{\parallel }^{\alpha \beta }\rightarrow \sigma _{\parallel }^{\alpha \beta
}/\Gamma _{\parallel }\;$. After some algebra, we can recast the structure
factors in much more compact form: 
\begin{eqnarray}
S^{++}({\bf k}) &=&\frac{{\bf k}\mbox{\boldmath $\sigma$}^{1}{\bf k}}{k^{2}}%
-({\bf k}\mbox{\boldmath $\sigma$}^{2}{\bf k})\frac{k^{2}}{%
k^{4}+4m^{2}k_{\parallel }^{2}}  \nonumber \\
\mbox{Re}\{S^{+-}({\bf k})\} &=&\frac{{\bf k}\mbox{\boldmath $\sigma$}^{3}%
{\bf k}}{k^{2}}-({\bf k}\mbox{\boldmath $\sigma$}^{4}{\bf k})\frac{k^{2}}{%
k^{4}+4m^{2}k_{\parallel }^{2}}  \label{scl_str} \\
\mbox{Im}\{S^{+-}({\bf k})\} &=&({\bf k}\mbox{\boldmath $\sigma$}^{5}{\bf k})%
\frac{\varepsilon \Gamma _{\parallel }^{\frac{1}{2}}\,k_{\parallel }}{%
k^{4}+4m^{2}k_{\parallel }^{2}}\;,  \nonumber
\end{eqnarray}
where $k=\mid {\bf k}\mid $ and 
\begin{eqnarray}
\mbox{\boldmath $\sigma$}^{1} &=&\frac{(1-\delta \tilde{\rho})^{2}}{(1-%
\tilde{\rho})((1-\delta \tilde{\rho})^{2}-\tilde{\rho}^{2})}\;%
\mbox{\boldmath $\sigma$}^{++}-\frac{(1-\tilde{\rho})\tilde{\rho}}{(1-\tilde{%
\rho})((1-\delta \tilde{\rho})^{2}-\tilde{\rho}^{2})}\;%
\mbox{\boldmath
$\sigma$}^{+-}  \nonumber \\
\mbox{\boldmath $\sigma$}^{2} &=&\left[ \frac{(1-\delta \tilde{\rho})^{2}(1-2%
\tilde{\rho})}{(1-\tilde{\rho})^{2}((1-\delta \tilde{\rho})^{2}-\tilde{\rho}%
^{2})}-1\right] \;\frac{(1-\tilde{\rho})}{(1-2\tilde{\rho})}\;%
\mbox{\boldmath
$\sigma$}^{++}  \nonumber \\
&-&\left[ \frac{(1-\delta \tilde{\rho})(1-2\tilde{\rho})}{(1-\tilde{\rho}%
)((1-\delta \tilde{\rho})^{2}-\tilde{\rho}^{2})}-1\right] \;\frac{\tilde{\rho%
}}{(1-2\tilde{\rho})}\;\mbox{\boldmath $\sigma$}^{+-}  \nonumber \\
\mbox{\boldmath $\sigma$}^{3} &=&\frac{(1-\tilde{\rho})\tilde{\rho}}{(1-%
\tilde{\rho})((1-\delta \tilde{\rho})^{2}-\tilde{\rho}^{2})}\;%
\mbox{\boldmath $\sigma$}^{++}-\frac{\tilde{\rho}^{2}}{(1-\tilde{\rho}%
)((1-\delta \tilde{\rho})^{2}-\tilde{\rho}^{2})}\;\mbox{\boldmath
$\sigma$}^{+-}  \label{sigmas} \\
\mbox{\boldmath $\sigma$}^{4} &=&\left[ \frac{(1-\delta \tilde{\rho})(1-2%
\tilde{\rho})}{(1-\tilde{\rho})((1-\delta \tilde{\rho})^{2}-\tilde{\rho}^{2})%
}+1\right] \;\frac{\tilde{\rho}}{(1-2\tilde{\rho})}\;%
\mbox{\boldmath
$\sigma$}^{++}  \nonumber \\
&-&\left[ \frac{\tilde{\rho}^{2}(1-2\tilde{\rho})}{(1-\tilde{\rho}%
)^{2}((1-\delta \tilde{\rho})^{2}-\tilde{\rho}^{2})}+1\right] \;\frac{(1-%
\tilde{\rho})}{(1-2\tilde{\rho})}\;\mbox{\boldmath $\sigma$}^{+-}  \nonumber
\\
\mbox{\boldmath $\sigma$}^{5} &=&\frac{\tilde{\rho}(2-(1+\delta )\tilde{\rho}%
)}{(1-\tilde{\rho})(1-2\tilde{\rho})}\;\mbox{\boldmath $\sigma$}^{++}-\frac{%
(1-\tilde{\rho})(1-\delta \tilde{\rho})+\tilde{\rho}^{2}}{(1-\tilde{\rho}%
)(1-2\tilde{\rho})}\;\mbox{\boldmath $\sigma$}^{+-}\;.  \nonumber
\end{eqnarray}
For the following, only two major properties of the $%
\mbox{\boldmath
$\sigma$}$'s, rather than their detailed expressions, are important. First,
they are all positive definite within the stability limit of the homogeneous
phase, and second, they are diagonal but generically not proportional to the
unit matrix, since they are related to the noise matrices. Now, referring
the details of the integrations to Appendix A, the transforms can be carried
out exactly. Writing $r\equiv |{\bf x}|$, $r_{\perp }\equiv |{\bf x}_{\perp
}|$ and $r_{\parallel }\equiv |x_{\parallel }|$, we define 
\begin{eqnarray}
E({\bf x}) &\equiv &\frac{\Gamma \left( \frac{d}{2}-1\right) }{4\pi ^{\frac{d%
}{2}}}\,\frac{1}{r^{d-2}}  \nonumber \\
F_{1}({\bf x}) &\equiv &\frac{\cosh (mx_{\parallel })}{(2\pi )^{\frac{d}{2}}}%
\,\left( \frac{m}{r}\right) ^{\frac{d-2}{2}}K_{\frac{d-2}{2}}(mr)
\label{EF_def} \\
F_{2}({\bf x}) &\equiv &\frac{\sinh (mx_{\parallel })}{(2\pi )^{\frac{d}{2}}}%
\,\left( \frac{m}{r}\right) ^{\frac{d-2}{2}}K_{\frac{d-2}{2}}(mr)\;, 
\nonumber
\end{eqnarray}
where $\Gamma (z)$ is the Gamma function and $K_{\nu }(z)$ is the modified
Bessel function. The correlations can then be expressed in terms of these
three functions: 
\begin{eqnarray}
G^{++}({\bf x}) &=&-%
\mbox{\boldmath $\nabla\sigma^{\mbox{\scriptsize
\unboldmath $1$}} \nabla$}\,E({\bf x})+%
\mbox{\boldmath
$\nabla\sigma^{\mbox{\scriptsize \unboldmath $2$}} \nabla$}\,F_{1}({\bf x})
\label{G++} \\
G_{e}^{+-}({\bf x}) &=&-%
\mbox{\boldmath $\nabla\sigma^{\mbox{\scriptsize
\unboldmath $3$}} \nabla$}\,E({\bf x})+%
\mbox{\boldmath
$\nabla\sigma^{\mbox{\scriptsize \unboldmath $4$}} \nabla$}\,F_{1}({\bf x})
\label{G1} \\
G_{o}^{+-}({\bf x}) &=&\frac{\varepsilon \Gamma _{\parallel }^{\frac{1}{2}}}{%
2m}\,%
\mbox{\boldmath $\nabla\sigma^{\mbox{\scriptsize \unboldmath $5$}}
\nabla$}\,F_{2}({\bf x})\;,  \label{G2}
\end{eqnarray}
where $G_{e,o}^{+-}$ are the parts of $G^{+-}$ even or odd in $x_{\parallel
} $, corresponding to the transforms of the real and imaginary parts of $%
S^{+-} $. The full correlation is, of course, $G^{+-}({\bf x})=G_{e}^{+-}(%
{\bf x})+G_{o}^{+-}({\bf x})$,$\;$reflecting the symmetry (\ref{g+-}) of the
system in the presence of the field.

Leaving the detailed asymptotic expansions of (\ref{G++}-\ref{G2}) to
Appendix B, we only indicate the main features here. Clearly, the
discontinuity singularities of the structure factors translate into
power-law decays of the correlation functions. In particular, the first
terms in (\ref{G++}) and (\ref{G1}) exhibit the well-known ${r^{-d}}$-decay 
\cite{powers} which is associated with the breaking of the FDT in the
presence of conservation laws \cite{ZHSL}. In contrast, a discontinuity
singularity in higher derivatives is not sufficient to produce the ${r^{-d}}$
power, as illustrated by the form of $G_{o}^{+-}$, (\ref{G2}). The second
terms in (\ref{G++}) and (\ref{G1}) and the only one in (\ref{G2}) are more
subtle, since the exponentials of the hyperbolic and Bessel functions
compete. Summarizing our results as $mr\rightarrow \infty $, we find for $%
r_{\perp }\neq 0$ : 
\begin{eqnarray}
G^{++}({\bf x}) &\propto &\frac{\sigma _{\parallel }^{1}-\sigma _{\perp }^{1}%
}{r^{d}}\left[ \frac{r_{\perp }^{2}-(d-1)r_{\parallel }^{2}}{r^{2}}\right]
+...  \nonumber \\
G^{+-}({\bf x})\simeq G_{e}^{+-}({\bf x}) &\propto &\frac{\sigma _{\parallel
}^{3}-\sigma _{\perp }^{3}}{r^{d}}\left[ \frac{r_{\perp
}^{2}-(d-1)r_{\parallel }^{2}}{r^{2}}\right] +...\;,  \label{gen_power} \\
G_{o}^{+-}({\bf x}) &\propto &...  \nonumber
\end{eqnarray}
where the $...$ represent exponentially decaying parts. Here, all three key
components of the characteristic non-equilibrium power law decays are
displayed, namely, the necessity of FDT violation, $\sigma _{\parallel
}/\Gamma _{\parallel }\neq \sigma _{\perp }/\Gamma _{\perp }$, the dipole
amplitude associated with strong anisotropy in the presence of a
conservation law, and the $r^{-d}$ itself. We emphasize again that the odd
part of the cross-correlations, $G_{o}^{+-}({\bf x})$, is purely
short-ranged.

Along the field direction, the behavior of the correlations is more complex.
Here, a novel power, $r_{\parallel }^{-(d+1)/2}$, emerges, which will
dominate over the ``FDT-violating'' $r^{-d}$, for all $d>1$. Thus, for $%
r_{\perp }=0$ we have: 
\begin{eqnarray}
G^{++}({\bf 0},x_{\parallel })\;,\;G_{e}^{+-}({\bf 0},x_{\parallel })
&\propto &\,-r_{\parallel }^{-\frac{d+1}{2}}+{\cal O}\left( \max \left\{
r_{\parallel }^{-d}\,,r_{\parallel }^{-\frac{d+3}{2}}\right\} \right) 
\nonumber \\
G_{o}^{+-}({\bf 0},x_{\parallel }) &\propto &-\mbox{sgn}(\varepsilon
x_{\parallel })\,r_{\parallel }^{-\frac{d+1}{2}}\,+{\cal O}\left(
r_{\parallel }^{-\frac{d+3}{2}}\right) \;.  \label{new power}
\end{eqnarray}
In the parentheses, we have indicated the {\em next}-leading term in the
asymptotic expansion of $K_{\frac{d-2}{2}}$. Surprisingly, in $d>3$ even
this power is still more relevant than the more familiar $r^{-d}$. We also
emphasize that in (\ref{new power}), all proportionality constants are
positive. Thus, the explicit factors of $(-1)$ carry information about the
structure of particle clusters in this model. In particular, the sign of $%
G_{o}^{+-}({\bf 0},x_{\parallel })$ shows that negative charges prefer to be
located `downfield', rather than `upfield', from positive ones, as a
precursor of the blocking transition. In conclusion, the spatial
correlations are dominated by the expected $r^{-d}$ power law, except along
the field, where a novel $r_{\parallel }^{-(d+1)/2}$ decay takes over.
Similar behavior is found if charge exchange is not allowed ($\gamma =0$) 
\cite{KSZ1}. Thus, this new power law appears to be a generic feature of
driven two-species models, associated with the excluded volume constraint
and the opposite bias. We should note, however, that it can only be
generated in the presence of at least one transverse dimension, i.e., in $%
d>1 $.

\subsection{The $\gamma=1$ case}

All expressions simplify considerably when we set $\gamma =1$, yet they
still capture the essence of this two species model, namely, the non-trivial
correlations between opposite charges: 
\begin{eqnarray}
S^{++}({\bf k}) &=&\frac{{\bf k}\mbox{\boldmath $\sigma$}^{++}{\bf k}}{k^{2}}
\nonumber \\
\mbox{Re}\{S^{+-}({\bf k})\} &=&({\bf k}\mbox{\boldmath $\sigma$}^{+-}{\bf k}%
)\frac{k^{2}}{k^{4}+4m^{2}k_{\parallel }^{2}}  \label{g1str} \\
\mbox{Im}\{S^{+-}({\bf k})\} &=&({\bf k}\mbox{\boldmath $\sigma$}^{+-}{\bf k}%
)\frac{-\mbox{sgn}(\varepsilon )\,2m\,k_{\parallel }}{k^{4}+4m^{2}k_{%
\parallel }^{2}}\;,  \nonumber
\end{eqnarray}
where now $4m^{2}=(1-2\bar{\rho})^{2}\varepsilon ^{2}\Gamma _{\parallel }$.
No instabilities can occur here: even for $\bar{\rho}=1/2$ where the
``mass'' $m^{2}$ vanishes, homogeneous configurations prevail since the
model reduces to a driven one-species model. For generic densities, we note
that the form of the $++$ structure factor is the same as in the one-species
model, due to the fact that $+$'s cannot distinguish between $-$'s and holes
at the microscopic level. The key question is, of course, whether the $%
\mbox{\boldmath $\sigma$}$'s (especially $\mbox{\boldmath $\sigma$}^{++}$)
are proportional to the unit matrix or not. Unfortunately, in the absence of
a renormalization group analysis we have to rely on simulations to answer
these questions. Based on the results of the previous subsection, it is
clear that only $S^{++}({\bf k})$ could possibly produce the $r^{-d}$ power
law. However, our simulation results indicate that the internal symmetry of
the system, at this particular value of $\gamma $, restores FDT for either
species, i.e., the first equation in (\ref{FDT}). This is entirely
consistent with the fact that the microscopic steady state distribution of
either species is uniform, as mentioned in \mbox{Section II}. Thus,
correlations will be short ranged, given by a $\delta $-function for
identical species and exponential decays for opposite charges, {\em except
in the field direction, between opposite species}, where: 
\begin{equation}
G^{+-}({\bf 0},x_{\parallel })\simeq 2\,\Theta (\varepsilon x_{\parallel
})\,\sigma _{\perp }^{+-}\frac{\sqrt{\frac{\pi }{2}}}{(2\pi )^{\frac{d}{2}}}%
\,\left\{ \frac{d-1}{2m}\left( \frac{m}{r_{\parallel }}\right) ^{\frac{d+1}{2%
}}+{\cal O}\left( \frac{1}{r_{\parallel }^{\frac{d+3}{2}}}\right) \right\}
\;.  \label{g1corr}
\end{equation}
Here $\Theta (x)$ is the step function and $\sigma _{\perp }^{+-}$ is always
negative. Thus, the novel $r_{\parallel }^{-(d+1)/2}$ power law, a key
feature of this two-species model, survives in the cross correlation, even
in this simplified case.

\subsection{Distribution of structure factors}

So far, we have focused entirely on the {\em averages} of density-density
operators. In this final section, we will construct the full {\em %
probability distributions} for these fluctuating quantities, i.e., $\frac{%
\chi ^{+}({\bf k},t)\chi ^{+}(-{\bf k},t)}{V}$, $\frac{\mbox{Re}[\chi ^{+}(%
{\bf k},t)\chi ^{-}(-{\bf k},t)]}{V}$ and $\frac{\mbox{Im}[\chi ^{+}({\bf k}%
,t)\chi ^{-}(-{\bf k},t)]}{V}$, following the method of ref. \cite{RZ}.
Representing these operators by $s^{++}$, $s_{r}^{+-}$ and $s_{i}^{+-}$, we
seek their marginal distributions, for each ${\bf k}$-vector separately: 
\begin{eqnarray}
P^{++}(s^{++};{\bf k}) &=&\left\langle \delta \left( \frac{\chi ^{+}({\bf k}%
,t){\chi ^{+}}^{*}({\bf k},t)}{V}-s^{++}\right) \right\rangle  \nonumber \\
P_{r}^{+-}(s_{r}^{+-};{\bf k}) &=&\left\langle \delta \left( \frac{\mbox{Re}%
[\chi ^{+}({\bf k},t){\chi ^{-}}^{*}({\bf k},t)]}{V}-s_{r}^{+-}\right)
\right\rangle  \label{prob_dist} \\
P_{i}^{+-}(s_{i}^{+-};{\bf k}) &=&\left\langle \delta \left( \frac{\mbox{Im}%
[\chi ^{+}({\bf k},t){\chi ^{-}}^{*}({\bf k},t)]}{V}-s_{i}^{+-}\right)
\right\rangle \;.  \nonumber
\end{eqnarray}
Here, we have used $\chi ^{\pm }(-{\bf k},t)={\chi ^{\pm }}^{*}({\bf k},t)$,
since the densities ${\chi ^{\pm }}({\bf r},t)$ are real. Also, we have
normalized by $V$ in order to obtain a well-defined thermodynamic limit,
noting that $(2\pi )^{d}\delta ({\bf k}\!=\!{\bf 0})=V$. In principle, these
distributions can be computed explicitly, by inserting the solution $\chi
^{\pm }({\bf k},t)$ of the Langevin equation (\ref{rsl}) into (\ref
{prob_dist}) and averaging over the noise, associated with (\ref{noise}).
However, given that these distributions are universal \cite{KSZ2}, depending
only on the linearity of the Langevin equation and the Gaussian nature of
the noise, rather than on the specific forms of diffusion and noise
matrices, (\ref{det_matr}) and (\ref{noisematr}), a detailed calculation is
not necessary. Instead, we can refer to the distributions for a simpler case 
\cite{KSZ2}, namely the model without charge exchange, since their forms
will be identical to the ones we are seeking here. However, some brief
comments are in order, to put the results into perspective. For technical
reasons, it is simpler to compute the characteristic functions (i.e.,
Fourier transforms) of (\ref{prob_dist}) first. Denoting these by $\tilde{P}%
^{\alpha \beta }(\Omega )$, we find that $\tilde{P}^{++}(\Omega )$ has a
single pole in the lower half $\Omega $-plane, so that the inverse transform
yields an exponential distribution for the non-negative variable $s^{++}$: 
\begin{equation}
P^{++}(s^{++};{\bf k})=\left\{ 
\begin{array}{cc}
\frac{1}{S^{++}({\bf k})}\;e^{-s^{++}/S^{++}({\bf k})} & \;\;\mbox{if}%
\;\;s^{++}\geq 0 \\ 
0 & \;\;\mbox{if}\;\;s^{++}<0
\end{array}
\right.  \label{pro++}
\end{equation}
Here, $S^{++}({\bf k})$ is just the {\em average }structure factor, and we
will refer to $1/S^{++}({\bf k})$ as the ``decay factor'' of the
exponential. In contrast, both $\tilde{P}_{r}^{+-}(\Omega )$ and $\tilde{P}%
_{i}^{+-}(\Omega )$ exhibit {\em two} poles, one ($\Omega _{-}$) being on
the negative, and one ($\Omega _{+}$) on the positive imaginary axis, 
\begin{equation}
\Omega _{\mp }=\frac{2i}{\Delta }\;\left( \mbox{Re}[S^{+-}({\bf k})]\mp 
\sqrt{\Delta +(\mbox{Re}[S^{+-}({\bf k})])^{2}}\right) \text{ \ }
\label{param}
\end{equation}
where $\Delta \equiv |S^{++}({\bf k})|^{2}-|S^{+-}({\bf k})|^{2}>0$. The
inverse transforms also result in exponential distributions, characterized
however by two distinct decay factors $|\Omega _{+}|$ and $|\Omega _{-}|$: 
\begin{equation}
P_{r}^{+-}(s_{r}^{+-};{\bf k})=\left\{ 
\begin{array}{cc}
\frac{1}{N}\;e^{-|\Omega _{-}|s_{r}^{+-}} & \;\;\mbox{if}\;\;s_{r}^{+-}\geq 0
\\ 
\frac{1}{N}\;e^{|\Omega _{+}|s_{r}^{+-}} & \;\;\mbox{if}\;\;s_{r}^{+-}<0
\end{array}
\right.  \label{pro+-}
\end{equation}
with $N=\sqrt{\Delta +(\mbox{Re}[S^{+-}({\bf k})])^{2}}$. The distribution
of $\frac{\mbox{Im}[\chi ^{+}(t){\chi ^{-}}^{*}(t)]}{V}$, i.e., $P_{i}^{+-}$%
, follows from $P_{r}^{+-}$ by just interchanging $\mbox{Re}[S^{+-}({\bf k}%
)] $ and $\mbox{Im}[S^{+-}({\bf k})]$ in Eqs. (\ref{param},\ref{pro+-}).

To summarize, all three distributions are asymmetric exponentials, with $%
P^{++}$ representing the most extreme case. Due to this structure, their
standard deviations are always {\em greater or equal} than the averages, so
that fluctuations will never be ``small'', in the usual sense \cite{KSZ2,RZ}.

\section{Discussion}

\begin{figure}[tbp]
\hspace*{2cm}
\epsfxsize=12cm
\epsfysize=12cm
\epsfbox{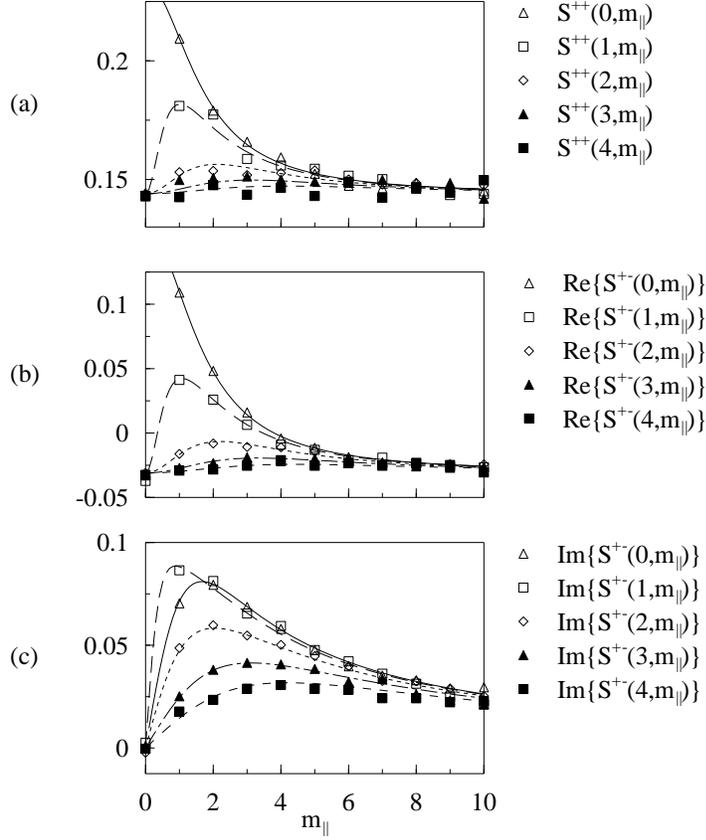}
\caption{Steady state structure factors (a) $S^{++}({\bf k})$ , (b) $
\mbox{Re}\{S^{+-}({\bf k})\}$ , (c) $\mbox{Im}\{S^{+-}({\bf k})\}$ for an $%
L=100$ system at $\gamma=0.02$ , $E=0.279$ and $\bar{\rho}=0.175$. Structure
factors are plotted against the integer $m_{\parallel}=\frac{k_{\parallel} L%
}{2\pi}$, while $m_{\perp}=\frac{k_{\perp} L}{2\pi}$ is taken as a
parameter. Lines are representing the fitted theoretical curves.}
\label{fig1}
\end{figure}

\begin{figure}[tbp]
\hspace*{2cm}
\epsfxsize=12cm
\epsfysize=12cm
\epsfbox{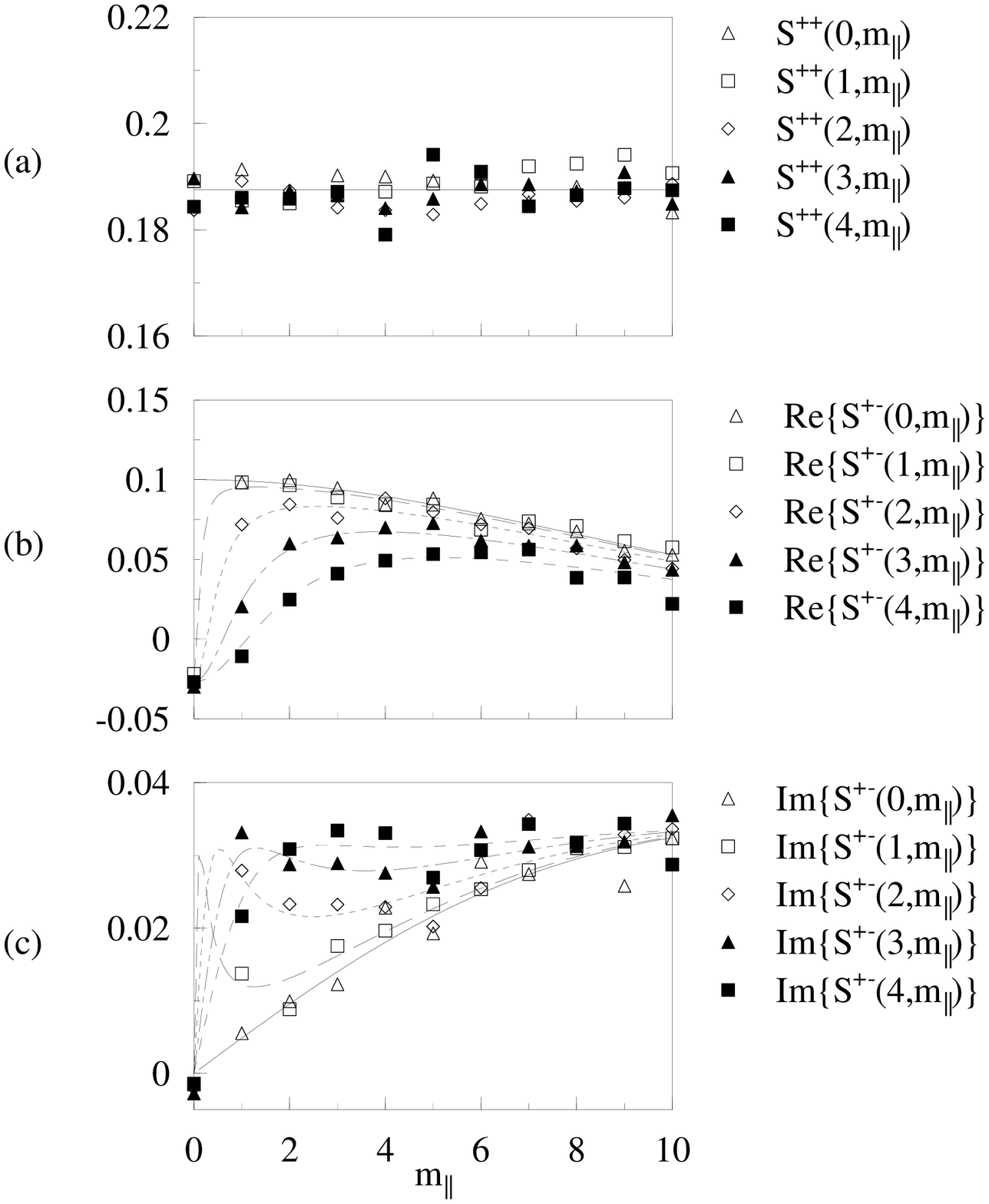}
\caption{Steady state structure factors (a) $S^{++}({\bf k})$ , (b) $
\mbox{Re}\{S^{+-}({\bf k})\}$ , (c) $\mbox{Im}\{S^{+-}({\bf k})\}$ for an $%
L=100$ system at $\gamma=1.00$ , $E=\infty$ and $\bar{\rho}=0.25$. Structure
factors are plotted against the integer $m_{\parallel}=\frac{k_{\parallel} L%
}{2\pi}$, while $m_{\perp}=\frac{k_{\perp} L}{2\pi}$ is taken as a
parameter. Lines are representing the fitted theoretical curves.}
\label{fig2}
\end{figure}

\begin{figure}[tbp]
\hspace*{2cm}
\epsfxsize=12cm
\epsfysize=12cm
\epsfbox{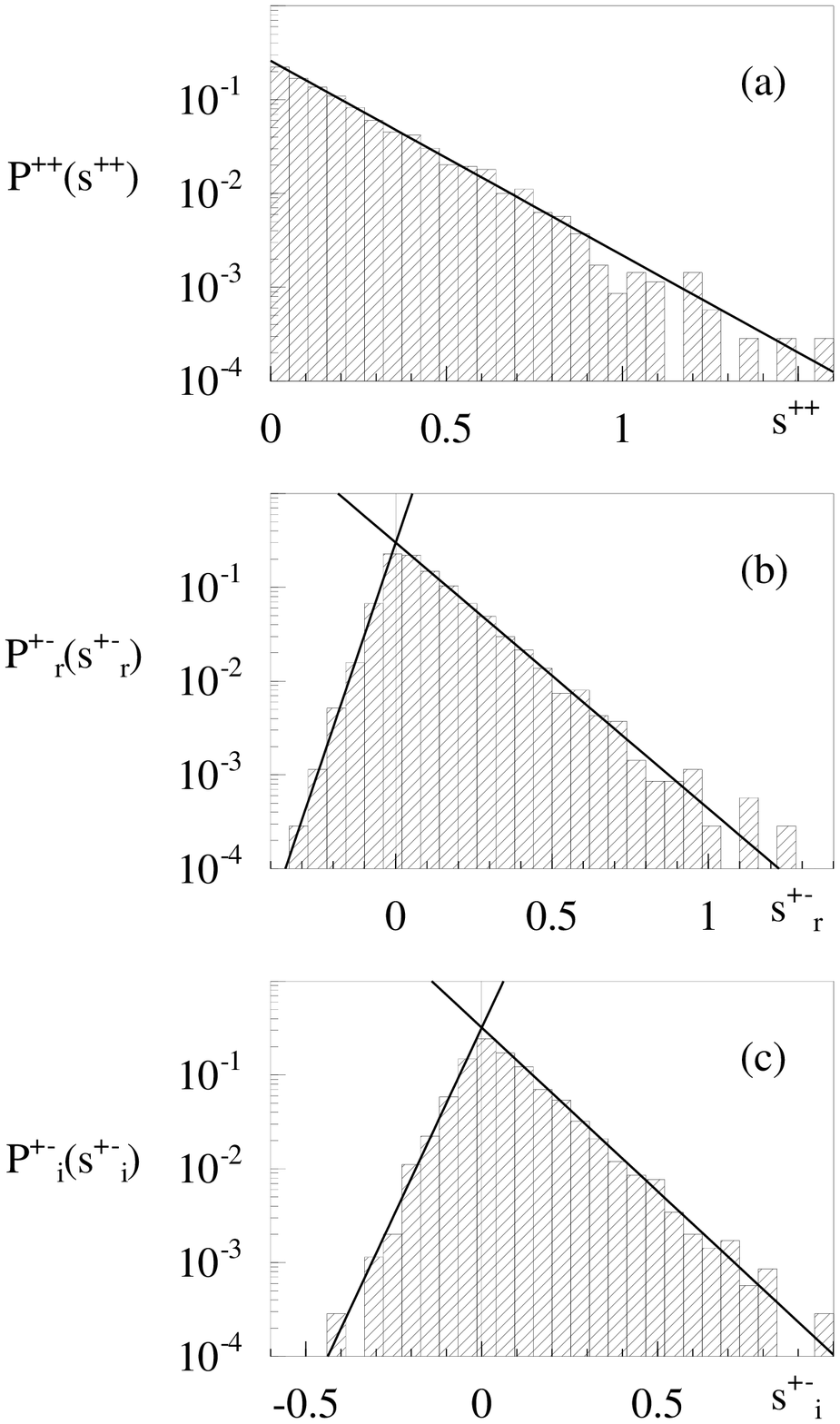}
\caption{Histograms representing the distributions of the ${\bf k}=\frac{%
2\pi }{L}(0,1)$ structure factors for (a) $\frac{ n_{{\bf k}}^{+} n_{-{\bf k}
}^{+} }{V} $, (b) $\frac{\mbox{Re}[n_{{\bf k}}^{+} n_{-{\bf k}}^{-}] }{V} $
and (c) $\frac{\mbox{Im}[n_{{\bf k}}^{+} n_{-{\bf k}}^{-}] }{V} $. $L=100$, $%
\gamma=0.02$, $E=0.279$ and $\bar{\rho}=0.175$. Theoretical distributions
(a) $P^{++}(s^{++};{\bf k})$, (b) $P^{+-}_{r}(s^{+-}_{r};{\bf k})$ and (c) $%
P^{+-}_{i}(s^{+-}_{i};{\bf k})$ are plotted with solid lines on the same
graphs. }
\label{fig3}
\end{figure}

\begin{figure}[tbp]
\hspace*{2cm}
\epsfxsize=12cm
\epsfysize=12cm
\epsfbox{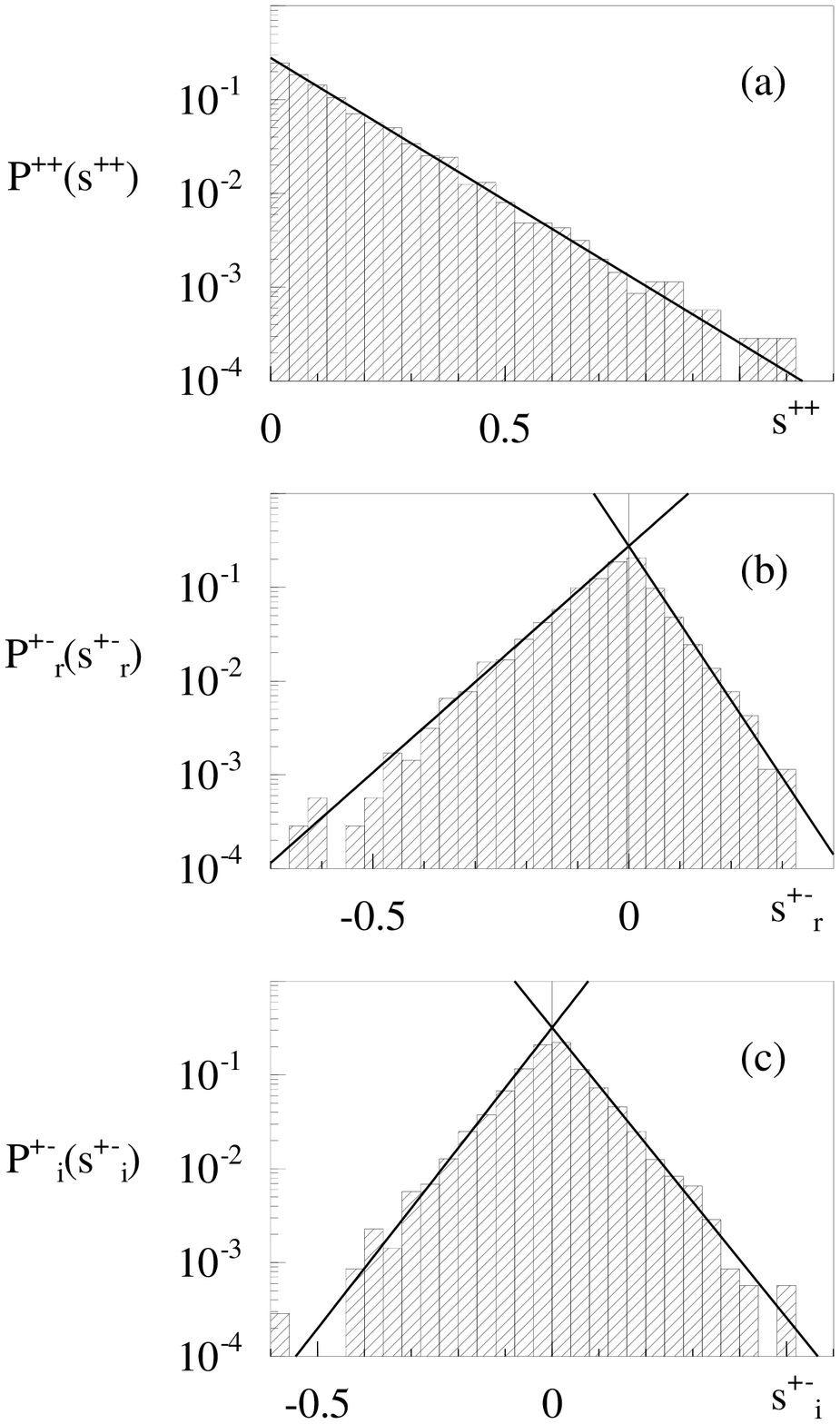}
\caption{Histograms representing the distributions of the ${\bf k}=\frac{%
2\pi }{L}(1,0)$ structure factors for (a) $\frac{ n_{{\bf k}}^{+} n_{-{\bf k}
}^{+} }{V} $, (b) $\frac{\mbox{Re}[n_{{\bf k}}^{+} n_{-{\bf k}}^{-}] }{V} $
and (c) $\frac{\mbox{Im}[n_{{\bf k}}^{+} n_{-{\bf k}}^{-}] }{V} $. $L=100$, $%
\gamma=0.02$, $E=0.279$ and $\bar{\rho}=0.175$. Theoretical distributions
(a) $P^{++}(s^{++};{\bf k})$, (b) $P^{+-}_{r}(s^{+-}_{r};{\bf k})$ and (c) $%
P^{+-}_{i}(s^{+-}_{i};{\bf k})$ are plotted with solid lines on the same
graphs. }
\label{fig4}
\end{figure}

Finally, let us turn to comparisons with simulation results. Typically, we
find that power law tails are much more difficult to observe than in the
single-species case \cite{powers}. Apparently, their amplitudes are rather
small, so that the data are obscured by either critical singularities or
finite size effects, depending on the points in the phase diagram which we
choose to investigate. Thus, we focus on the structure factors. Using a
standard least-square routine, we fitted our analytical results (eqs. (\ref
{scl_str}) and (\ref{g1str}) before rescaling) to our simulation data. The
fit was done simultaneously for the three $S$'s using the smallest $5\times
11$ non-zero ${\bf k}$ vectors. The agreement is quite good, especially
considering that the theoretical results are based on a linearized Langevin
equation, but we note the following: for the $\gamma =0.02$ case (Fig. \ref
{fig1}), despite being in the homogeneous phase, the system was relatively
close to the continuous transition \cite{KSZ}, with 
\mbox{$m \sim 4 \times
10^{-2}$} corresponding to a correlation length \mbox{$\xi \sim 25$} in
units of the lattice constant. In particular, ``longitudinal'' parameters,
such as $\Gamma _{\parallel }$ and the $\sigma _{\parallel }$'s, seem to
suffer considerable renormalizations here. On the other hand, the
``transverse'' parameters, $\Gamma _{\bot }$ and $\sigma _{\bot }$, appear
to obey Eqn. (\ref{FDT}). In that sense, the FDT is satisfied {\em within
the transverse subspace}. To illustrate this feature, we combine (\ref{FDT}%
), written for the transverse parameters, with the explicit form of the
structure factors (\ref{exp_str}) for $k_{\parallel }=0$. This yields the
exact ``finite size'' amplitudes, completely independent of ${\bf k}_{\perp }
$: $S^{++}({\bf k}_{\perp },0)=\bar{\rho}(1-\bar{\rho})$ and $S^{+-}({\bf k}%
_{\perp },0)=-\bar{\rho}^{2}$, in perfect agreement with the simulations. In
the {\em full} $d$-dimensional space, however, the FDT is of course
violated: As a result of the coarse-graining effect in the field direction,
we generically found 
\mbox{$\sigma_{\parallel}/\Gamma_{\parallel} \neq 
\sigma_{\perp}/\Gamma_{\perp}$} for the rescaled ``noise'' matrices. In
particular we had 
\mbox{$\sigma^{1}_{\parallel}/\Gamma_{\parallel} = 
0.833\,\sigma^{1}_{\perp}/\Gamma_{\perp}$}, predicting the typical
FDT-violating power law.

For $\gamma =1.00$ (Fig. \ref{fig2}), $S^{++}({\bf k})$ is completely flat,
as we expected, indicating that 
\mbox{$\sigma^{++}_{\parallel}/\Gamma_{\parallel}=
\sigma^{++}_{\perp}/\Gamma_{\perp}$}. Moreover, the value of this constant
is just $\bar{\rho}(1-\bar{\rho})$, again consistent with (\ref{FDT}). In
contrast, $S^{+-}({\bf k})$ clearly exhibits the structure of Eqs. (\ref
{g1str}). Here the system is far from transitions ($\xi \sim 3$), so that
critical fluctuations are completely avoided. Consequently, using 
\mbox{$m=\frac{1}{2}(1-2\bar{\rho})|\varepsilon|
\Gamma_{\parallel}^{\frac{1}{2}}$} with the mean-field parameters produces a
``mass'' closely matching the one obtained from the fit.

Now, we turn to a comparison of the analytical results for the structure
factor distributions with the simulations, summarized in Figs. \ref{fig3}
and \ref{fig4}, for the two smallest wave vectors, respectively. The control
parameters were the same as those of Fig. \ref{fig1}. Again, the agreement
between our Gaussian theory and the data is quite impresssive. The `$++$'
histograms show simple exponential decay \cite{RZ}, while the `$+-$'
histograms clearly represent asymmetric exponential distributions. To test
the theoretical prediction, namely, that the slopes of the histograms are
determined entirely by the structure factor averages, we simply measured the
latter, i.e., $S^{++}$ , $\mbox{Re}S^{+-}$ and $\mbox{Im}S^{+-}$. We then
inserted the {\em measured} averages into the {\em theoretical} relations
for the decay factors. Clearly, the `$++$' case is particularly simple since
the decay factor is just the inverse of $S^{++}$ itself. For the two '$+-$'
distributions, the decay factors $|\Omega _{\mp }|$ , given by (\ref{param}%
), are considerably less trivial, but the agreement is nevertheless
remarkable. Here, renormalizations can obviously also occur, but can be
absorbed into the effective parameters of the theory, leaving the {\em form }%
of the structure factor distributions invariant. Moreover, they are
automatically captured by the {\em measured} structure factors, so that they
do not spoil the agreement between data and theory here. However, we must
avoid critical fluctuations since these fall out of the scope of a linear
theory.

In summary, using both simulations and analytic techniques, we have examined
the structure factors in a simple model of biased diffusion of two species.
We calculated the corresponding spatial correlations, finding not only the
expected power law decay, $r^{-d}$, typical for non-equilibrium steady
states of conserved systems in the presence of strong anisotropy, but also a 
{\em novel} power, $r_{\parallel }^{-(d+1)/2}$, for correlations along the
bias, characteristic for two-species models. We also investigated the full
distribution functions for the structure factors, being universal asymmetric
distributions. The general agreement between simulations and a Gaussian
field theory is surprisingly good, while we await a renormalization group
analysis of the continuum theory of the model in order to make more detailed
comparisons closer to the continuous transition.

\section*{Acknowledgements}

We thank R.K.P. Zia, Z. Toroczkai and S. Sandow for many stimulating
discussions. This research is supported in part by grants from the National
Science Foundation through the Division of Materials Research .

\appendix

\section{Momentum-space Integrals for the Correlation Functions}

From Eqn. (\ref{scl_str}) we see that we need three basic type of integrals.
Although the first one is well known, we list it for completeness: 
\begin{equation}
E({\bf x})\equiv \int \frac{d^{d}k}{(2\pi )^{d}}\,\frac{e^{i{\bf kx}}}{k^{2}}%
=\frac{\Gamma \left( \frac{d}{2}-1\right) }{4\pi ^{\frac{d}{2}}}\,\frac{1}{%
r^{d-2}}\;,  \label{E_def}
\end{equation}
Then, if $\mbox{\boldmath $\sigma$}$ is diagonal and isotropic in the $d-1$
dimensional transverse subspace but not a multiple of the unit matrix, it is
easy to compute 
\begin{equation}
\mbox{\boldmath $\nabla\sigma\nabla$}\,E({\bf x})=-\sigma _{\perp }\delta (%
{\bf x})-(\sigma _{\parallel }-\sigma _{\perp })\,\frac{\Gamma \left( \frac{d%
}{2}\right) }{2\pi ^{\frac{d}{2}}}\frac{r_{\perp }^{2}-(d-1)r_{\parallel
}^{2}}{r^{d+2}}\;.  \label{DE}
\end{equation}
Next, we will outline a formal way to obtain the other two required momentum
integrals. For a more rigorous treatment see \cite{PhD}. We define $F_{1}$
and $F_{2}$ as follows: 
\begin{eqnarray}
F_{1}({\bf x}) &\equiv &\int \frac{d^{d}k}{(2\pi )^{d}}\,\frac{e^{i{\bf kx}%
}\,k^{2}}{k^{4}+4m^{2}k_{\parallel }^{2}}  \nonumber \\
F_{2}({\bf x}) &\equiv &\int \frac{d^{d}k}{(2\pi )^{d}}\,\frac{e^{i{\bf kx}%
}\,(-2m)ik_{\parallel }}{k^{4}+4m^{2}k_{\parallel }^{2}}\;.  \label{F1F2_def}
\end{eqnarray}
It is then helpful to realize that the integrands, without the exponential
factor, are simply the convolutions of two functions, i.e.: 
\begin{eqnarray}
\frac{k^{2}}{k^{4}+4m^{2}k_{\parallel }^{2}} &=&\int \frac{d^{d}k^{\prime }}{%
(2\pi )^{d}}\,F({\bf k}^{\prime })C({\bf k}-{\bf k}^{\prime })  \nonumber \\
\frac{-2mik_{\parallel }}{k^{4}+4m^{2}k_{\parallel }^{2}} &=&\int \frac{%
d^{d}k^{\prime }}{(2\pi )^{d}}\,F({\bf k}^{\prime })S({\bf k}-{\bf k}%
^{\prime })\;,  \label{conv}
\end{eqnarray}
where 
\begin{eqnarray}
F({\bf k}) &=&\frac{1}{k^{2}+m^{2}}  \nonumber \\
C({\bf k}) &=&\frac{(2\pi )^{d}}{2}\delta ({\bf k}_{\perp })\,\left[ \delta
(k_{\parallel }+im)+\delta (k_{\parallel }-im)\right]  \label{ktr} \\
S({\bf k}) &=&\frac{(2\pi )^{d}}{2}\delta ({\bf k}_{\perp })\,\left[ \delta
(k_{\parallel }+im)-\delta (k_{\parallel }-im)\right] \;.  \nonumber
\end{eqnarray}
The $\delta $-functions with complex arguments should only be understood in
an operational sense. The Fourier inverse transforms of these functions are
easily found: 
\begin{eqnarray}
F({\bf x}) &\equiv &F(r)=\frac{1}{(2\pi )^{\frac{d}{2}}}\,\left( \frac{m}{r}%
\right) ^{\frac{d-2}{2}}K_{\frac{d-2}{2}}(mr)  \nonumber \\
C({\bf x}) &=&\cosh (mx_{\parallel })  \label{rsp} \\
S({\bf x}) &=&\sinh (mx_{\parallel })\;.  \nonumber
\end{eqnarray}
Thus, using the convolution theorem, we trivially get 
\begin{eqnarray}
F_{1}({\bf x}) &=&\cosh (mx_{\parallel })F(r)  \nonumber \\
F_{2}({\bf x}) &=&\sinh (mx_{\parallel })F(r)\;.  \label{F1F2}
\end{eqnarray}
Note that $F(r)$ is the solution of 
\begin{equation}
\left( -\mbox{\boldmath $\nabla$}^{2}+m^{2}\right) F(r)\,=\,\delta ({\bf x}%
)\;.  \label{basic}
\end{equation}
Then using some algebra and (\ref{basic}), we can translate $%
\mbox{\boldmath
$\nabla\sigma\nabla$}$ into differentiation with respect to $x_{\parallel }$%
: 
\begin{eqnarray}
\mbox{\boldmath $\nabla\sigma\nabla$}\,F_{1}({\bf x}) &=&-\sigma _{\perp
}\delta ({\bf x})+\sigma _{\perp }2m\,\partial _{\parallel }\left\{ \sinh
(mx_{\parallel })F(r)\right\} +(\sigma _{\parallel }-\sigma _{\perp
})\,\partial _{\parallel }^{2}\left\{ \cosh (mx_{\parallel })F(r)\right\} 
\nonumber \\
&=&-\sigma _{\perp }\delta ({\bf x})+\sigma _{\perp }2m\,\partial
_{\parallel }\,F_{2}({\bf x})+(\sigma _{\parallel }-\sigma _{\perp
})\,\partial _{\parallel }^{2}\,F_{1}({\bf x})  \nonumber \\
\mbox{\boldmath $\nabla\sigma\nabla$}\,F_{2}({\bf x}) &=&\sigma _{\perp
}2m\,\partial _{\parallel }\left\{ \cosh (mx_{\parallel })F(r)\right\}
+(\sigma _{\parallel }-\sigma _{\perp })\,\partial _{\parallel }^{2}\left\{
\sinh (mx_{\parallel })F(r)\right\}  \label{DFS} \\
&=&\sigma _{\perp }2m\,\partial _{\parallel }\,F_{1}({\bf x})+(\sigma
_{\parallel }-\sigma _{\perp })\,\partial _{\parallel }^{2}\,F_{2}({\bf x}%
)\;.  \nonumber
\end{eqnarray}
These forms are particularly useful when we calculate the corresponding
long-distance behavior.

\section{Long Distance Asymptotic Behavior of the Correlation Functions}

To obtain the long-distance behavior for $%
\mbox{\boldmath
$\nabla\sigma\nabla$}\,E({\bf x})$, we just have to omit the first term in (%
\ref{DE}), which is a $\delta $-function: 
\begin{equation}
\left. \mbox{\boldmath $\nabla\sigma\nabla$}\,E({\bf x})\right| _{{\bf x}%
\neq {\bf 0}}=-(\sigma _{\parallel }-\sigma _{\perp })\,\frac{\Gamma \left( 
\frac{d}{2}\right) }{2\pi ^{\frac{d}{2}}}\frac{r_{\perp
}^{2}-(d-1)r_{\parallel }^{2}}{r^{d+2}}\;.  \label{LDE}
\end{equation}
This is the typical ``FDT-violating'' power law, provided that $%
\mbox{\boldmath $\sigma$}$ is not a simple multiple of the unit matrix.
Otherwise, the amplitude of this term would be zero.

Using the ``large $z$'' asymptotic expansion of the modified Besssel
function \cite{RG} 
\begin{equation}
K_{\nu }(z)\simeq \sqrt{\frac{\pi }{2z}}\,e^{-z}\left\{ 1+\left( \nu ^{2}-%
\frac{1}{4}\right) \,\frac{1}{2z}+{\cal O}\left( \frac{1}{z^{2}}\right)
\right\} \;,  \label{BAE}
\end{equation}
we can obtain the long-distance behavior for $%
\mbox{\boldmath
$\nabla\sigma\nabla$}\,F_{1}({\bf x})$ and $%
\mbox{\boldmath
$\nabla\sigma\nabla$}\,F_{2}({\bf x})$ as \mbox{$m=\mbox{const.}>0$} and %
\mbox{$r \rightarrow \infty$}. Due to the strong anisotropies in these
functions, we consider three different scenarios:

\noindent (i.) $r_{\parallel }=0\;,\;\;\;r_{\perp }\rightarrow \infty \;:$ 
\newline
Combining (\ref{DFS}) and the asymptotic form of $F(r)$ we find: 
\begin{eqnarray}
\left. \partial _{\parallel }\,F_{2}({\bf x})\right| _{x_{\parallel }=0}
&=&\left. mF(r)\right| _{x_{\parallel }=0}\simeq \frac{\sqrt{\frac{\pi }{2}}%
}{(2\pi )^{\frac{d}{2}}}\,e^{-mr_{\perp }}\left\{ \left( \frac{m}{r_{\perp }}%
\right) ^{\frac{d-1}{2}}+{\cal O}\left( \frac{1}{r_{\perp }^{\frac{d+1}{2}}}%
\right) \right\}  \nonumber \\
\left. \partial _{\parallel }^{2}\,F_{1}({\bf x})\right| _{x_{\parallel }=0}
&=&\left. m^{2}F(r)\right| _{x_{\parallel }=0}+\frac{1}{r}\left. \frac{%
\partial F(r)}{\partial r}\right| _{x_{\parallel }=0} \\
&\simeq &m\frac{\sqrt{\frac{\pi }{2}}}{(2\pi )^{\frac{d}{2}}}\,e^{-mr_{\perp
}}\left\{ \left( \frac{m}{r_{\perp }}\right) ^{\frac{d-1}{2}}+{\cal O}\left( 
\frac{1}{r_{\perp }^{\frac{d+1}{2}}}\right) \right\} \;,  \nonumber
\end{eqnarray}
while $\partial _{\parallel }\,F_{1}({\bf x})$ and $\partial _{\parallel
}^{2}\,F_{2}({\bf x})$ are simply zero at $x_{\parallel }=0$, since they are
odd functions of $x_{\parallel }$. Thus, finally we have 
\begin{eqnarray}
\left. \mbox{\boldmath $\nabla\sigma\nabla$}\,F_{1}({\bf x})\right|
_{x_{\parallel }=0} &\simeq &(\sigma _{\perp }+\sigma _{\parallel })\,m\,%
\frac{\sqrt{\frac{\pi }{2}}}{(2\pi )^{\frac{d}{2}}}\,e^{-mr_{\perp }}\left\{
\left( \frac{m}{r_{\perp }}\right) ^{\frac{d-1}{2}}+{\cal O}\left( \frac{1}{%
r_{\perp }^{\frac{d+1}{2}}}\right) \right\}  \nonumber \\
\left. \mbox{\boldmath $\nabla\sigma\nabla$}\,F_{2}({\bf x})\right|
_{x_{\parallel }=0} &=&0  \label{FY0}
\end{eqnarray}

\noindent (ii.) $r_{\parallel }\rightarrow \infty \;,\;\;\;r_{\perp }\neq
0\;:$ \newline
In addition to using the asymptotic form of $F(r)$, we can now also write %
\mbox{$\cosh(mx_{\parallel}) \simeq \frac{1}{2} e^{mr_{\parallel}}$} and 
\mbox{$\sinh(mx_{\parallel}) \simeq \mbox{sgn}(x_{\parallel}) \frac{1}{2}
e^{mr_{\parallel}}$}. In the following, we will keep the second leading
power in $1/r$ in order to simplify the discussion of case (iii). We find 
\begin{eqnarray}
\partial _{\parallel }\,F_{2}({\bf x}) &\simeq &\frac{\sqrt{\frac{\pi }{2}}}{%
(2\pi )^{\frac{d}{2}}}\,\frac{e^{mr_{\parallel }}}{2}\,e^{-mr}\left\{ \left(
1-\frac{r_{\parallel }}{r}\right) \left( \frac{m}{r}\right) ^{\frac{d-1}{2}%
}\right.  \nonumber \\
&+&\left. \frac{d-1}{8m^{2}}\left( (d-3)-(d+1)\frac{r_{\parallel }}{r}%
\right) \left( \frac{m}{r}\right) ^{\frac{d+1}{2}}+{\cal O}\left( \frac{1}{%
r^{\frac{d+3}{2}}}\right) \right\}  \nonumber \\
\partial _{\parallel }^{2}\,F_{1}({\bf x}) &\simeq &m\frac{\sqrt{\frac{\pi }{%
2}}}{(2\pi )^{\frac{d}{2}}}\,\frac{e^{mr_{\parallel }}}{2}\,e^{-mr}\left\{
\left( 1-\frac{r_{\parallel }}{r}\right) ^{2}\left( \frac{m}{r}\right) ^{%
\frac{d-1}{2}}\right.  \label{F12XY} \\
&+&\left. \frac{d+1}{8m^{2}}\left( (d-5)-2(d-1)\frac{r_{\parallel }}{r}+(d+3)%
\frac{r_{\parallel }^{2}}{r^{2}}\right) \left( \frac{m}{r}\right) ^{\frac{d+1%
}{2}}+{\cal O}\left( \frac{1}{r^{\frac{d+3}{2}}}\right) \right\}  \nonumber
\\
\partial _{\parallel }\,F_{1}({\bf x}) &\simeq &\mbox{sgn}(x_{\parallel
})\,\partial _{\parallel }\,F_{2}({\bf x})  \nonumber \\
\partial _{\parallel }^{2}\,F_{2}({\bf x}) &\simeq &\mbox{sgn}(x_{\parallel
})\,\partial _{\parallel }^{2}\,F_{1}({\bf x})  \nonumber
\end{eqnarray}
Thus, for $r_{\perp }\neq 0$ we have in leading order: 
\begin{eqnarray}
\mbox{\boldmath $\nabla\sigma\nabla$}\,F_{1}({\bf x}) &\simeq &m\frac{\sqrt{%
\frac{\pi }{2}}}{(2\pi )^{\frac{d}{2}}}\,\frac{e^{mr_{\parallel }}}{2}%
\,e^{-mr}\left\{ \left( (\sigma _{\perp }+\sigma _{\parallel })-2\sigma
_{\parallel }\frac{r_{\parallel }}{r}+(\sigma _{\parallel }-\sigma _{\perp })%
\frac{r_{\parallel }^{2}}{r^{2}}\right) \left( \frac{m}{r}\right) ^{\frac{d-1%
}{2}}\right.  \nonumber \\
&+&\left. {\cal O}\left( \frac{1}{r^{\frac{d+1}{2}}}\right) \right\}
\label{FXY} \\
\mbox{\boldmath $\nabla\sigma\nabla$}\,F_{2}({\bf x}) &\simeq &\mbox{sgn}%
(x_{\parallel })\,\mbox{\boldmath $\nabla\sigma\nabla$}\,F_{1}({\bf x})\;. 
\nonumber
\end{eqnarray}

\noindent (iii.) $r_{\parallel }\rightarrow \infty \;,\;\;\;r_{\perp }=0\;:$ 
\newline
Note that (\ref{F12XY}) was obtained exploiting only \mbox{$r_{\parallel}%
\rightarrow \infty$}. Setting \mbox{$r_{\perp}=0$} has two important
consequences: since now \mbox{$r=r_{\parallel}$}, the exponential decays
cancel and, further, the amplitude of the $(1/r)^{\frac{d-1}{2}}$ term will
vanish. Invoking the next-to-leading terms in (\ref{F12XY}) yields: 
\begin{eqnarray}
\left. \mbox{\boldmath $\nabla\sigma\nabla$}\,F_{1}({\bf x})\right|
_{r_{\perp }=0} &\simeq &-\sigma _{\perp }\frac{\sqrt{\frac{\pi }{2}}}{(2\pi
)^{\frac{d}{2}}}\,\left\{ \frac{d-1}{2m}\left( \frac{m}{r_{\parallel }}%
\right) ^{\frac{d+1}{2}}+{\cal O}\left( \frac{1}{r_{\parallel }^{\frac{d+3}{2%
}}}\right) \right\}  \nonumber \\
\left. \mbox{\boldmath $\nabla\sigma\nabla$}\,F_{2}({\bf x})\right|
_{r_{\perp }=0} &\simeq &\mbox{sgn}(x_{\parallel })\left. 
\mbox{\boldmath
$\nabla\sigma\nabla$}\,F_{1}({\bf x})\right| _{r_{\perp }=0}\;.  \label{FX0}
\end{eqnarray}
which are the desired results.

\newpage


\end{document}